\documentclass[sigconf]{acmart}
\usepackage{setspace}
\setcopyright{none}
\renewcommand\footnotetextcopyrightpermission[1]{}

\usepackage{graphicx}
\usepackage{booktabs}
\usepackage{adjustbox}
\usepackage{tcolorbox}
\usepackage{url}
\author{Daniel Silva}
\email{danielanderson.ti@gmail.com}
\affiliation{%
  \institution{Federal University of Campina Grande - UFCG}
  \country{Brazil}
}

\author{Renan Alves}
\email{jose.pereira@embedded.ufcg.edu.br}
\affiliation{%
  \institution{Federal University of Campina Grande - UFCG}
  \country{Brazil}
}

\author{Emanuel Dantas Filho}
\email{emanuel.filho@jaboatao.ifpe.edu.br}
\affiliation{%
  \institution{Federal Institute of Pernambuco - IFPE}
  \country{Brazil}
}

\author{Ademar Sousa Neto}
\email{ademar.1409131@educar.rn.gov.br}
\affiliation{%
  \institution{Federal University of Campina Grande - UFCG}
  \country{Brazil}
}

\author{Mirko Perkusich}
\email{mirko@virtus.ufcg.edu.br}
\affiliation{%
  \institution{Federal University of Campina Grande - UFCG}
  \country{Brazil}
}

\author{Danyllo Wagner Albuquerque}
\email{danyllo.albuquerque@ifpb.edu.br}
\affiliation{%
  \institution{Instituto Federal da Paraíba - IFPB}
  \country{Brazil}
}

\author{Kyller Gorgônio}
\email{kyller@virtus.ufcg.edu.br}
\affiliation{%
  \institution{Federal University of Campina Grande - UFCG}
  \country{Brazil}
}

\author{Angelo Perkusich}
\email{perkusic@virtus.ufcg.edu.br}
\affiliation{%
  \institution{Federal University of Campina Grande - UFCG}
  \country{Brazil}
}

\tcbset{
  rqbox/.style={
    colback=gray!5,
    colframe=gray!70,
    boxrule=0.5pt,
    arc=3pt,
    left=3pt,
    right=3pt,
    top=3pt,
    bottom=3pt,
    fontupper=\small

  }
}

\title{Structural Validation of LLM-Generated Microservice Decompositions Using Source-Code Dependencies}

\begin{abstract}
Decomposing monolithic systems into microservices is a key activity in software modernization. Although Large Language Models (LLMs) can generate semantically plausible decompositions from textual requirements, it remains unclear whether these proposals preserve the structural dependencies implemented in the source code. This paper evaluates the structural adherence of microservice decompositions generated by OpenAI o3 for the PetClinic and Bookstore systems. We propose an automated validation pipeline based on static dependency analysis and compare zero-shot and few-shot prompting using dependency preservation (TPD) and dependency violation (TVD) metrics. A robustness analysis was conducted to control for differences in class-to-service mapping coverage. After normalization, both prompting strategies produced equivalent structural adherence, achieving TPD values of 68.0\% (PetClinic) and 83.3\% (Bookstore). The findings demonstrate that structural evaluations of LLM-generated decompositions should explicitly control for mapping coverage, as apparent differences between prompting strategies may otherwise reflect methodological bias rather than genuine architectural quality.
\end{abstract}


\begin{document}
\maketitle


\section{Introduction}
\label{sec:intro}

Modernizing legacy monolithic systems to microservice architectures remains one of the main challenges of contemporary Software Engineering. Although this migration promises gains in scalability, flexibility, and maintainability, its success depends on identifying architectural boundaries that preserve high cohesion and low coupling. In practical terms, this means grouping functionalities that belong to the same business context and avoiding separating components with strong structural dependencies in the source code. Without such care, the migration can result in a distributed architecture that is more complex, more costly, and potentially more fragile than the monolithic system itself.

In recent years, Large Language Models (LLMs) have played an increasing role in Software Engineering activities, including code generation, documentation, refactoring, and architectural design support. Several studies have investigated using these models to generate microservice decompositions directly from textual requirements, obtaining promising results in service identification, with \(F_1\)-score values close to 0.97~\cite{Albuquerque2026, Pereira2025}. These results suggest that LLMs can understand domain semantics and propose conceptually coherent architectures, making them natural candidates to support software modernization.

However, when used solely with textual requirement descriptions, these approaches lack access to the structure actually implemented in the source code. As a consequence, a semantically plausible decomposition may separate components that exhibit strong structural coupling, creating inter-service dependencies that will require additional communication, refactoring, or integration mechanisms during migration. In other words, semantic quality does not necessarily imply structural feasibility. This distinction becomes critical when proposed architectures support real modernization decisions, where technical constraints and existing dependencies play a fundamental role.

Recent research seeks to reduce this limitation by incorporating structural information into the decomposition process itself. Approaches such as MonoEmbed~\cite{Sellami2026} and MicroDec~\cite{Alsayed2024b}, for example, use source code analysis to produce more consistent architectures. However, these approaches focus on generating decompositions from code, while a complementary problem remains relatively unexplored: how to structurally validate architectures already produced by LLMs from textual requirements. Likewise, few experimental protocols exist for comparing prompting strategies while controlling for methodological biases.

From the perspective of Intelligent Software Engineering, this gap shifts the focus from merely generating architectures to validating architectural decisions produced by AI systems. Existing LLM-based approaches can suggest semantically plausible decompositions, but there is limited evidence on whether these decompositions respect the monolith's existing dependency structure. Therefore, instead of assuming that a conceptually coherent architecture is structurally feasible, it becomes necessary to confront AI-produced recommendations with objective evidence extracted from the source code. This validation step increases transparency, helps architects understand the structural implications of generated decompositions, and contributes to making the use of LLMs in modernization tasks more auditable and reproducible.

In this context, this study proposes an automated and reproducible structural validation pipeline for LLM-generated microservice decompositions. The approach was applied to decompositions produced by OpenAI o3 for the PetClinic and Bookstore monolithic systems, using zero-shot and few-shot strategies. In addition to comparing strategies, a robustness analysis was conducted to control for differences in class-to-service mapping coverage, revealing that mapping coverage may bias structural comparisons between prompting strategies if left uncontrolled. 

This study makes three main contributions. First, it identifies and mitigates a methodological bias caused by differences in class-to-service mapping coverage, showing that apparent differences between prompting strategies may be artifacts of incomplete mappings rather than genuine improvements in decomposition quality. Second, it proposes an automated and reproducible pipeline for structurally validating LLM-generated microservice decompositions using static analysis and dependency graphs. Finally, it introduces an experimental protocol based on the TVD and TPD metrics to quantify the structural adherence between generated decompositions and the original source code.

The remainder of this paper is organized as follows. Section~\ref{sec:metodologia} presents the study methodology, including the experimental design, the analyzed systems, the proposed pipeline, and the metrics used. Section~\ref{sec:resultados} presents the experimental results and the robustness analysis. Section~\ref{sec:discussao} discusses the main findings in light of the literature. Section~\ref{sec:implicacoes} presents the implications for research and industry. Section~\ref{sec:ameacas} discusses threats to validity, and Section~\ref{sec:conclusao} concludes the work and presents directions for future research.

\section{Background and Research Gap}
\label{sec:background}

Decomposing monoliths into microservices is a central challenge in the modernization of legacy systems. Classic works such as Newman~\cite{Newman2015} and Fowler~\cite{Fowler2014} emphasize the importance of well-defined boundaries, low coupling, and high cohesion. However, identifying such boundaries requires understanding both domain responsibilities and the structural dependencies existing in the source code. Recent reviews, such as that of Abgaz et al.~\cite{Abgaz2023}, show that the area still lacks public datasets, unified metrics, and comparable protocols to evaluate monolith decompositions.

With the advancement of LLMs, studies have begun to investigate the generation of microservice architectures from textual requirements. Albuquerque et al.~\cite{Albuquerque2026} and Pereira et al.~\cite{Pereira2025,Pereira2026a} show that models such as OpenAI o3 and Claude 3.5 Sonnet can identify services with high conceptual accuracy. Pereira et al.~\cite{Pereira2026b} complement this line of work by characterizing structural properties of generated architectures, observing differences between prompting strategies, such as the tendency of few-shot prompting to produce more fine-grained decompositions, with a larger number of microservices and more detailed descriptions. However, these works predominantly evaluate the semantic coherence of decompositions, without systematically verifying whether the proposed boundaries preserve real source-code dependencies.

Another line of research seeks to incorporate structural information directly into the decomposition process. Approaches such as MonoEmbed~\cite{Sellami2026}, MicroDec~\cite{Alsayed2024b}, MicroRec~\cite{Alsayed2024_MicroRec}, and graph-based techniques~\cite{Sooksatra2024} explore static analysis, embeddings, dependency graphs, and AI models to support microservice identification. These works reinforce the importance of combining semantic and structural signals. However, they focus on generating decompositions from code, whereas the present study investigates a complementary problem: how to structurally validate decompositions already produced by LLMs from textual requirements.

The present study positions itself in this gap by proposing a structural validation pipeline attachable to any LLM-based architectural synthesis process. Instead of modifying the model or the prompt, the approach confronts the generated decompositions with a dependency graph extracted from the monolith, quantifying dependency violations and preservations through the TVD and TPD metrics.

\section{Study Settings}
\label{sec:metodologia}

\begin{figure*}[!ht]
    \centering
    \includegraphics[width=0.95\textwidth]{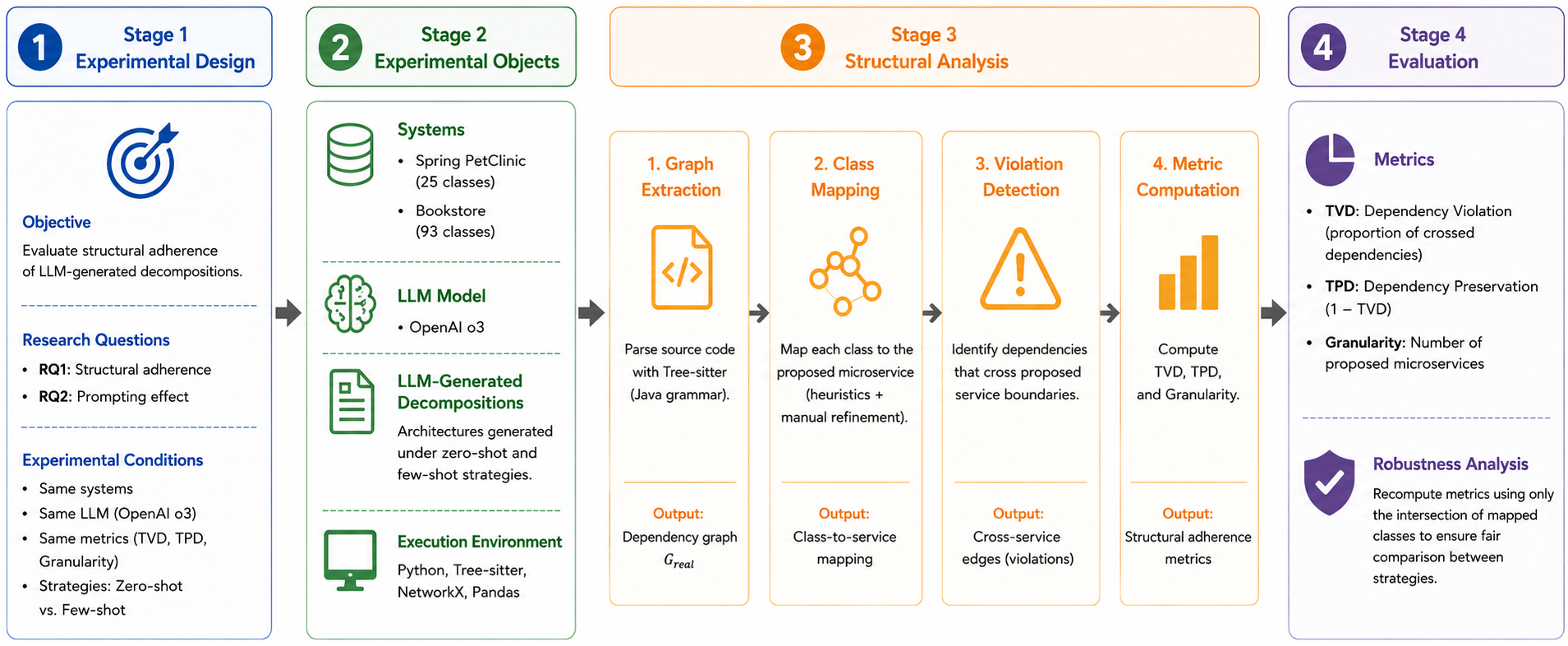}
    \caption{Overview of the proposed research methodology.}
    \label{fig:methodology}
\end{figure*}

Figure~\ref{fig:methodology} summarizes the research methodology adopted in this study. The process is organized into four stages: (i) experimental design, (ii) definition of the experimental objects, (iii) execution of the structural validation pipeline, and (iv) evaluation of the generated decompositions through structural metrics and robustness analysis.

\textbf{Stage 1: Experimental Design}. The objective of this study is to investigate whether the prompting strategy influences the structural adherence of LLM-generated microservice decompositions to the original source code. To isolate this variable, both experimental conditions employ the same monolithic systems (PetClinic and Bookstore), the same LLM (OpenAI o3), and the same evaluation metrics (TVD, TPD, and granularity), differing only in the prompting strategy (zero-shot and few-shot).

The study is guided by the following research questions (RQs):

\begin{description}
    \item[RQ1.] What is the structural adherence of LLM-generated microservice decompositions with respect to the actual source-code dependencies?
    
    \item[RQ2.] How do zero-shot and few-shot prompting affect the structural adherence of the generated decompositions?
\end{description}

\textit{RQ1} establishes a baseline by quantifying the structural adherence of LLM-generated decompositions with respect to the original source-code dependencies. This is important because semantically plausible decompositions may still conflict with the structural organization of the implemented system. \textit{RQ2} investigates whether the adopted prompting strategy (zero-shot or few-shot) influences this structural adherence. Understanding this effect is important for determining whether improvements reported for prompting strategies also translate into structurally more consistent decompositions.

\vspace{0.4em}
\textbf{Stage 2: Experimental Objects}. The evaluation considers two open-source Java monolithic systems selected because they differ in size and architectural complexity, allowing the proposed pipeline to be assessed on both a small benchmark and a larger application.

\begin{itemize}
\item \textit{PetClinic}: 25 domain classes, widely used as a benchmark for monolith decomposition research.

\item \textit{Bookstore}: 93 domain classes, representing a larger e-commerce application with multiple business domains and authentication mechanisms.
\end{itemize}

The analyzed decompositions were previously generated by OpenAI o3 under zero-shot and few-shot prompting strategies in the studies of Albuquerque et al.~\cite{Albuquerque2026} and Pereira et al.~\cite{Pereira2025,Pereira2026a}. For each system, dependency graphs were extracted through static analysis using Tree-sitter and the Java grammar, providing the structural baseline for the evaluation. The structural validation pipeline was implemented in Python using Tree-sitter, NetworkX, and Pandas and executed under the same environment for all experimental conditions.

\vspace{0.4em}
\textbf{Stage 3: Structural Analysis}. The proposed pipeline automatically validates LLM-generated microservice decompositions by comparing their service boundaries against structural dependencies extracted from the original monolith. The process comprises four sequential steps:

\begin{enumerate}
    \item \textit{Static extraction.} The source code is parsed with Tree-sitter and the Java grammar to construct the dependency graph $G_{\mathit{real}}$, capturing inheritance, implementations, dependency injection, repository usage, and JPA relationships.
    
    \item \textit{Class-to-service mapping.} Each source-code class is associated with one of the generated microservices using heuristics based on package and class names, followed by manual refinement whenever ambiguities are identified.
    
    \item \textit{Violation detection.} The dependency graph is analyzed to identify edges crossing service boundaries, corresponding to structural violations introduced by the proposed decomposition.
    
    \item \textit{Metric computation.} The detected violations are used to compute TVD, TPD, and decomposition granularity, providing a quantitative assessment of structural adherence.
\end{enumerate}

The proposed pipeline produces a complete set of artifacts that make the structural validation process auditable and reproducible. These include dependency graphs (GraphML), class-to-service mappings (CSV), structured violation reports, and the computed TVD and TPD metrics. Together, these artifacts allow every reported result to be traced back to the corresponding source-code dependencies, facilitating independent verification and reproduction of the experiments.

The pipeline also highlighted two methodological aspects relevant to reproducibility. First, class-to-service mapping coverage directly influences the computed metrics and must therefore be controlled when comparing decompositions. Second, although dependency extraction is largely automated, manual auditing remained necessary for a small number of framework-mediated relationships (e.g., \texttt{Vet~$\rightarrow$~Specialty}), reducing potential extraction errors. 

\vspace{0.4em}
\textbf{Stage 4: Evaluation}. The generated decompositions are evaluated using three complementary metrics that quantify their structural adherence to the original monolith.

\begin{itemize}
    \item \textit{TVD (Total Violated Dependencies):} proportion of extracted dependencies that cross proposed service boundaries. Higher values indicate greater structural divergence from the original monolith.
    
    \item \textit{TPD (Total Preserved Dependencies):} complement of TVD (\(TPD = 1 - TVD\)). It represents the proportion of extracted dependencies preserved within the same proposed microservice.
    
    \item \textit{Granularity:} total number of proposed microservices, reported as a descriptive characteristic of the generated decompositions rather than as a structural adherence metric.
\end{itemize}

Formally, let \(E\) denote the set of extracted dependencies and \(s(c)\) the function that assigns class \(c\) to a microservice. A dependency \((c_i,c_j)\in E\) is considered preserved if \(s(c_i)=s(c_j)\); otherwise, it is classified as a structural violation. Thus,

\[
TVD=
\frac{|\{(c_i,c_j)\in E \mid s(c_i)\neq s(c_j)\}|}{|E|}
\qquad
TPD=1-TVD
\]

TVD and TPD quantify structural adherence rather than overall architectural quality. A dependency crossing a service boundary is not necessarily undesirable, as it may result from legitimate architectural decisions such as refactoring, API-based communication, or domain-driven redesign. Therefore, these metrics should be interpreted as indicators of the potential structural adaptation effort required by a proposed decomposition, complementing semantic evaluations of service boundaries.

\section{Results}
\label{sec:resultados}

This section presents the results obtained from applying the pipeline to the PetClinic and Bookstore systems. Initially, the structural adherence of the model-generated decompositions is analyzed (RQ1). Next, the impact of the prompting strategy is investigated (RQ2), including a robustness analysis aimed at controlling coverage differences between the produced mappings. 

\subsection{Structural Adherence of Generated Decompositions (RQ1)}
\label{sec:resultados-rq1}

Table~\ref{tab:resultados-brutos} presents the preliminary structural adherence results, computed before normalizing the class-to-service mapping coverage.

\begin{table}[h]
  \centering
  \caption{Raw results -- PetClinic and Bookstore with OpenAI o3.}
  \label{tab:resultados-brutos}
  \small
  \setlength{\tabcolsep}{4pt}
  \begin{tabular}{lcccc}
    \toprule
    \textbf{Metric} & \multicolumn{2}{c}{\textbf{PetClinic}} & \multicolumn{2}{c}{\textbf{Bookstore}} \\
    \cmidrule(lr){2-3} \cmidrule(lr){4-5}
    & Zero-shot & Few-shot & Zero-shot & Few-shot \\
    \midrule
    Analyzed edges & 40 & 40 & 203 & 230 \\
    Structural violations & 9 & 8 & 34 & 52 \\
    TVD (Violation) & 22.5\% & 20.0\% & 16.8\% & 22.6\% \\
    TPD (Preservation) & 77.5\% & 80.0\% & 83.2\% & 77.4\% \\
    Granularity (\# serv.) & 8 & 8 & 6 & 7 \\
    \bottomrule
  \end{tabular}
\end{table}

The raw results show distinct behaviors between the two systems. In PetClinic, the difference between strategies was small: few-shot produced one fewer structural violation than zero-shot (TVD of 20.0\% versus 22.5\%). In Bookstore, however, zero-shot appeared to outperform few-shot, exhibiting 34 structural violations compared with 52. Because all experimental factors other than the prompting strategy were kept constant, these differences suggested the presence of an additional source of variation. This observation motivated a detailed inspection of the class-to-service mappings generated by each strategy.

Inspection of the mappings revealed that the observed differences did not stem from the structural organization of the decompositions, but from the class-to-service mapping coverage available for auditing. Because some monolith classes had not been assigned to a microservice in certain previously published decompositions, the metrics were initially computed over different sets of source-code dependencies. Under these conditions, the direct comparison between strategies introduces a methodological bias, as each is evaluated over a distinct universe of structural dependencies.

In Bookstore, zero-shot did not map the 13 classes responsible for JWT-based authentication, whereas few-shot grouped them into a dedicated authentication service. Since this subsystem concentrates several structural dependencies, excluding these classes removed 18 potential violations from the zero-shot evaluation, creating an apparent advantage that was unrelated to the prompting strategy itself. A similar phenomenon occurred in PetClinic, where the class \texttt{PetClinicApplication} was mapped only by zero-shot, introducing one exclusive structural violation.

To eliminate this bias, the metrics were recomputed considering only the intersection of the classes mapped by both strategies, ensuring that the comparison was performed over exactly the same set of source-code dependencies. The normalized results are presented in Table~\ref{tab:resultados-normalizados}.

\begin{table}[!ht]
  \centering
  \caption{Results normalized by the intersection of class-to-service mappings.}
  \label{tab:resultados-normalizados}
  \small
  \setlength{\tabcolsep}{4pt}
  \begin{tabular}{lcccc}
    \toprule
    \textbf{Metric} & \multicolumn{2}{c}{\textbf{PetClinic}} & \multicolumn{2}{c}{\textbf{Bookstore}} \\
    \cmidrule(lr){2-3} \cmidrule(lr){4-5}
    & Zero-shot & Few-shot & Zero-shot & Few-shot \\
    \midrule
    Classes in intersection & 15 & 15 & 68 & 68 \\
    Analyzed edges & 25 & 25 & 203 & 203 \\
    Structural violations & \textbf{8} & \textbf{8} & \textbf{34} & \textbf{34} \\
    \textbf{TVD} & \textbf{32.0\%} & \textbf{32.0\%} & \textbf{16.7\%} & \textbf{16.7\%} \\
    \textbf{TPD} & \textbf{68.0\%} & \textbf{68.0\%} & \textbf{83.3\%} & \textbf{83.3\%} \\
    \bottomrule
  \end{tabular}
\end{table}

After normalization, the results became identical in both systems. In PetClinic, both strategies achieved a TVD of 32.0\% and a TPD of 68.0\%. In Bookstore, both obtained a TVD of 16.7\% and a TPD of 83.3\%. These findings indicate that the differences observed in the preliminary analysis were entirely explained by differences in mapping coverage. Once this source of bias was removed, zero-shot and few-shot exhibited equivalent structural adherence in both evaluated systems. More broadly, this result demonstrates that comparing structural metrics over different mapping coverages can lead to misleading conclusions regarding the effectiveness of prompting strategies.

\begin{tcolorbox}[rqbox]
\textbf{Answer to RQ1 (Structural Adherence).}
After normalization, both strategies preserved a substantial but incomplete portion of the source code's dependencies. The preservation of 68.0\% of dependencies in PetClinic and 83.3\% in Bookstore indicates that the generated decompositions captured part of the systems' structural organization, while still leaving a non-negligible set of dependencies crossing the proposed service boundaries.
\end{tcolorbox}

\subsection{Impact of Prompting Strategy on Adherence (RQ2)}
\label{sec:resultados-rq2}

Although the normalized results indicate no quantitative differences between zero-shot and few-shot prompting, the violation reports generated by the pipeline provide qualitative insights into the types of structural inconsistencies introduced by both strategies. The analysis of structural violations identified by the pipeline revealed two recurring patterns. The first corresponds to the separation of entities related by JPA associations into distinct microservices. The second involves the displacement of controllers to services other than those containing the repositories used by those classes.

\sloppy
A representative example occurs in PetClinic, where \texttt{PetController} directly depends on \texttt{OwnerRepository}. In the analyzed decomposition, these classes were assigned to \texttt{Client Service} and \texttt{Pet Service}, respectively, causing this dependency to cross the boundary between microservices. This example illustrates a recurring structural violation pattern observed across the analyzed decompositions, rather than a limitation of a particular prompting strategy.

It is important to highlight that a structural violation does not, by itself, imply an incorrect architecture. It merely indicates that an existing dependency in the monolith would cross the proposed service boundary, requiring additional communication mechanisms, refactoring, or architectural reorganization during the migration process, which can increase coupling and consequently application latency. Thus, TVD should be interpreted as an indicator of potential structural adaptation effort, not as an absolute measure of architectural quality.

After controlling for the effect of mapping coverage, no differences were observed between the zero-shot and few-shot strategies in either of the analyzed systems. Moreover, both strategies exhibited the same categories of structural violations, suggesting that the remaining inconsistencies are primarily associated with the limited structural information available from textual requirements rather than with the prompting strategy itself.

Regarding decomposition granularity, the differences between prompting strategies were also limited (8 vs. 8 services in PetClinic and 6 vs. 7 in Bookstore), indicating that both produced architectures of comparable size. Therefore, the equivalent structural adherence observed after normalization cannot be attributed to substantial differences in decomposition granularity.

\begin{tcolorbox}[rqbox]
\textbf{Answer to RQ2 (Impact of Prompting Strategy).}
After normalization by the intersection of mapped classes, zero-shot and few-shot produced structurally equivalent results in both analyzed systems. The differences observed in the raw metrics are fully explained by differences in class-to-service mapping coverage, with no evidence of superiority of one strategy over the other when both are compared under the same conditions.
\end{tcolorbox}

\section{Discussion}
\label{sec:discussao}

The results provide two main insights into the structural validation of LLM-generated microservice decompositions. First, the normalized TPD values (68.0\% for PetClinic and 83.3\% for Bookstore) indicate that LLMs preserve a substantial portion of the monolith's structural organization. However, the remaining dependency violations show that semantic plausibility alone does not guarantee structural adherence. These findings are consistent with previous studies demonstrating that LLMs can identify conceptually coherent services from textual requirements~\cite{Albuquerque2026,Pereira2025,Pereira2026a}, while reinforcing evidence that structural information is essential for producing technically feasible decompositions~\cite{Mansour2025}.

From an architectural perspective, TVD should be interpreted as an estimate of the structural adaptation effort required during migration rather than as an indicator of poor architectural quality. Dependencies crossing service boundaries identify locations where architects may need to introduce APIs, refactor responsibilities, or redesign service boundaries. Consequently, the proposed metrics complement semantic evaluations by highlighting where migration effort is likely to concentrate.

Our findings also support hybrid approaches that combine semantic and structural information, such as MonoEmbed~\cite{Sellami2026} and MicroDec~\cite{Alsayed2024b}. The recurring violation patterns observed in this study, including the separation of strongly coupled entities and controllers from their associated repositories, suggest that these inconsistencies arise because textual requirements do not capture implementation dependencies. This observation reinforces the argument of Sellami~\cite{Sellami2026_Tese} that combining multiple sources of architectural evidence is essential for structurally consistent decompositions.

One of the most important contributions of this work is methodological. The apparent superiority of one prompting strategy disappeared after controlling for class-to-service mapping coverage. After normalization, zero-shot and few-shot produced identical structural adherence metrics, demonstrating that comparisons based solely on raw metrics can be misleading when decompositions cover different subsets of the source code.

More broadly, the results suggest that the observed limitations are more closely related to the absence of structural information in the input than to the prompting strategy itself. Under the evaluated conditions, zero-shot and few-shot converged to equivalent structural behavior once mapping coverage was controlled. Therefore, future advances in LLM-assisted architectural synthesis are likely to depend less on prompt engineering and more on integrating semantic information from requirements with structural evidence extracted from source code.

\section{Implications}
\label{sec:implicacoes}

The findings of this study extend beyond the evaluation of two prompting strategies. They provide methodological guidance for future empirical studies on LLM-assisted software architecture and practical insights for practitioners adopting LLMs in software modernization workflows. The implications span both research and industrial practice, highlighting how structural validation can complement semantic assessments and support more reliable architectural decision-making.

\vspace{0.4em}

\textbf{For research}, the results suggest that structural adherence metrics such as TVD and TPD should complement semantic evaluation metrics in benchmarks for LLM-based architectural synthesis. Evaluating only the conceptual quality of generated decompositions may overlook important structural inconsistencies with the underlying implementation. Furthermore, our robustness analysis demonstrates that comparisons between prompting strategies should explicitly report class-to-service mapping coverage and normalize evaluations whenever coverage differs, since raw structural metrics can be substantially biased by differences in the analyzed class set. Finally, the need to manually correct framework-mediated dependencies (e.g., \texttt{Vet~→~Specialty}) indicates opportunities for future research on hybrid extraction techniques that combine static analysis, dynamic analysis, and framework-aware dependency recovery.

\vspace{0.4em}

\textbf{For industrial practice}, the proposed pipeline can serve as an automated architectural auditing mechanism before migration decisions are made. Rather than treating LLM-generated decompositions as final architectural designs, practitioners can use the reported structural violations to identify coupling hotspots, estimate migration effort, and prioritize manual architectural review. The results also suggest that structural validation can be naturally incorporated into modernization workflows—for example, during architecture reviews or migration planning—providing objective evidence about the structural consequences of candidate decompositions while allowing architects to focus their effort on the relatively small subset of critical dependency violations.

\section{Threats to Validity}
\label{sec:ameacas}

The findings should be interpreted considering the limitations inherent to the adopted methodology and experimental setting.

\vspace{0.4em}

\textbf{External validity.} The evaluation considered only two Java monolithic systems and a single LLM (OpenAI o3). Although the selected systems differ in size and architectural complexity, the findings may not generalize to other programming languages, application domains, industrial-scale systems, alternative architectural styles, or newer language models. Replications involving additional systems and LLMs are therefore needed to assess the broader applicability of the proposed pipeline and the observed methodological findings.

\vspace{0.4em}

\textbf{Internal validity.} The structural metrics depend on correctly mapping source-code classes to the generated microservices. Although automated heuristics and manual inspection were employed, ambiguous mappings may affect the computed TVD and TPD values. Furthermore, the evaluated decompositions were obtained from previously published studies rather than generated within the present experimental protocol, which may introduce characteristics specific to those datasets. Differences in mapping coverage can also bias comparisons between prompting strategies; this threat was mitigated through the robustness analysis based on the intersection of mapped classes.

\vspace{0.4em}

\textbf{Construct validity.} TVD and TPD quantify structural adherence rather than overall architectural quality. Dependencies crossing service boundaries may represent acceptable design decisions after refactoring or API-based communication. Moreover, the proposed metrics do not capture other architectural quality attributes, such as scalability, maintainability, performance, fault isolation, or domain alignment. Finally, the static analysis does not identify runtime behaviors such as reflection, dynamic proxies, or framework-mediated interactions, although manual inspection was used to reduce extraction errors.

\vspace{0.4em}

\textbf{Conclusion validity.} The evaluation relied on deterministic metrics and descriptive comparisons without statistical hypothesis testing. Consequently, the observed equivalence between zero-shot and few-shot prompting should be interpreted only for the analyzed systems and experimental conditions. Replications involving additional systems, prompting strategies, and LLMs are necessary to strengthen the generality of the conclusions.

\vspace{0.4em}

\section{Final Remarks}
\label{sec:conclusao}

This study investigated the structural adherence of LLM-generated microservice decompositions with respect to the original source-code dependencies. To this end, we proposed an automated and reproducible validation pipeline based on static dependency analysis, class-to-service mapping, and structural adherence metrics. The methodology was applied to decompositions generated by OpenAI o3 for the PetClinic and Bookstore systems under zero-shot and few-shot prompting strategies, enabling an objective comparison between the generated architectures and the dependency structure of the original monoliths.

The results showed that the generated decompositions preserved a substantial portion of the original dependency structure, achieving TPD values of 68.0\% for PetClinic and 83.3\% for Bookstore, while still introducing structural violations requiring architectural adaptation during migration. More importantly, the apparent superiority observed between prompting strategies disappeared after controlling for class-to-service mapping coverage. After normalization, zero-shot and few-shot exhibited equivalent structural adherence, demonstrating that differences initially attributed to prompting strategy were actually caused by methodological bias.

These findings have implications for both research and practice. For research, they highlight the importance of complementing semantic evaluations with structural adherence metrics and explicitly controlling mapping coverage when comparing LLM-based architectural synthesis approaches. For practice, the proposed pipeline provides an auditable mechanism for validating AI-generated decompositions before migration decisions are made, helping architects identify structural inconsistencies and estimate adaptation effort based on objective evidence extracted from the source code.

Future work includes extending the evaluation to additional benchmark systems, such as MediaStore and TeaStore, and investigating other LLMs, prompting strategies, and decomposition approaches. Another promising direction is integrating the proposed validation pipeline into iterative architecture generation workflows, allowing structural feedback to guide successive generations toward more reliable and explainable AI-assisted software modernization.

\section*{Artifact Availability}
\label{sec:artifacts}

An anonymized replication package is available at \url{https://doi.org/10.5281/zenodo.20709439}. It contains the complete analysis pipeline, input and intermediate datasets, generated results, and instructions required to reproduce the experiments.

\section*{Generative AI Use}

Generative AI tools were used to support language editing and figure preparation. All scientific content was reviewed and approved by the authors, who take full responsibility for the manuscript.


\begin{thebibliography}{19}

\bibitem{Albuquerque2026}
Anonymous.
\textit{Anonymized}.
Submitted for review, 2026.
Available at: \url{https://submission-2026.com/paper01}

\bibitem{Pereira2025}
Pereira, Jose Renan A. \textit{et al.} Toward Generating Microservice Architectures from Textual Requirements with Large Language Models.
In: \textit{Anais do XIX Simpósio Brasileiro de Componentes, Arquiteturas e Reutilização de Software (SBCARS)}, p.~79--89, Porto Alegre, RS, 2025.

\bibitem{Sellami2026}
Sellami, Khaled; Saied, Mohamed Aymen. MonoEmbed: Enhancing LLM representations for monolith to microservices decomposition through contrastive learning.
\textit{Empirical Software Engineering}, v.~31, n.~11, 2026.

\bibitem{Alsayed2024b}
Alsayed, Ahmed Saeed; Dam, Hoa Khanh; Nguyen, Chau. MicroDec: Leveraging Large Language Models for Microservice Decomposition.
\textit{Journal of Systems and Software}, v.~203, 111622, 2024.

\bibitem{Pereira2026a}
Pereira, José Renan A. \textit{et al.} Do LLMs Agree on Microservice Decompositions? A Multi--Model Study from Textual Requirements.
In: \textit{The 41st ACM/SIGAPP Symposium on Applied Computing (SAC '26)}, Thessaloniki, Grécia, 2026.

\bibitem{Newman2015}
Newman, Sam. \textit{Building Microservices: Designing Fine-Grained Systems}. Sebastopol, CA: O'Reilly Media, 2015.

\bibitem{Fowler2014}
Lewis, James; Fowler, Martin. Microservices: a definition of this new architectural term. ThoughtWorks, 2014. \url{https://martinfowler.com/articles/microservices.html}

\bibitem{Krause2020}
Krause, Alexander \textit{et al.} Microservice Decomposition via Static and Dynamic Analysis of the Monolith.
In: \textit{2020 IEEE International Conference on Software Architecture Companion (ICSA-C)}, p.~9--16, 2020.

\bibitem{Matias2020}
Matias, Tiago \textit{et al.} Determining Microservice Boundaries: A Case Study Using Static and Dynamic Software Analysis.
In: \textit{European Conference on Software Architecture (ECSA)}, p.~315--332, 2020.

\bibitem{Abgaz2023}
Abgaz, Yalemisew \textit{et al.} Decomposition of Monolith Applications into Microservices Architectures: A Systematic Review.
\textit{IEEE Transactions on Software Engineering}, v.~49, n.~8, p.~4213--4242, 2023.

\bibitem{Bucaioni2025}
Bucaioni, Alessio \textit{et al.} Artificial Intelligence for Software Architecture: Literature Review and the Road Ahead.
In: \textit{Proceedings of the ACM International Conference on the Foundations of Software Engineering (FSE '25)}, ACM, 2025.

\bibitem{Pereira2026b}
Anonymous.
\textit{Anonymized}.
Submitted for review, 2026.
Available at: \url{https://submission-2026.com/paper02}

\bibitem{Mansour2025}
Ait Mansour, Nassima \textit{et al.} Semantic Approaches to Microservice Identification: A Systematic Literature Review.
\textit{IEEE Access}, v.~13, 2025.

\bibitem{Sooksatra2024}
Sooksatra, Korn \textit{et al.} Using Static Analysis to Aid Monolith to Microservice System Transformation: Tuning Fuzzy c-Means in a VAE-Based GNN Approach.
In: \textit{39th IEEE/ACM International Conference on Automated Software Engineering Workshops (ASEW '24)}, p.~43--53, 2024.

\bibitem{Saucedo2024_Variability}
Saucedo, Ana Martínez \textit{et al.} On the Variability of Microservice Decompositions: A Data-Driven Analysis.
In: \textit{2024 L Latin American Computer Conference (CLEI)}, p.~1--9, 2024.

\bibitem{Saucedo2024_Exploring}
Saucedo, Ana Martínez \textit{et al.} Exploring Alternative Microservice Decompositions using Data-driven Techniques and LLMs.
\textit{CLEI Electronic Journal}, v.~28, n.~3, paper 6, 2025.

\bibitem{Sellami2026_Tese}
Sellami, Khaled. Automated Migration of Monolithic Systems into Microservices-based Architecture.
217 f. Tese (Doutorado em Informática) -- Université Laval, Québec, Canadá, 2026.

\bibitem{Weerasinghe2026}
Weerasinghe, Mineth \textit{et al.} From Monolith to Microservices: A Comparative Evaluation of Decomposition Frameworks.
\textit{arXiv preprint arXiv:2601.23141}, 2026.

\bibitem{Alsayed2024_MicroRec}
Alsayed, Ahmed Saeed; Dam, Hoa Khanh; Nguyen, Chau. MicroRec: Leveraging Large Language Models for Microservice Recommendation.
In: \textit{21st International Conference on Mining Software Repositories (MSR '24)}, ACM, 2024.

\end{thebibliography}
\end{document}